\pdfoutput=1
\documentclass[11pt]{article}
\usepackage{times}
\usepackage{tensor}
\usepackage[pdftex]{graphicx}
\usepackage{natbib}
\usepackage{amsthm}
\usepackage{tensind}
\tensordelimiter{?}
\newcommand{\n}{\noindent}
\begin{document}

\title{Flavour-Oscillation Clocks and the Geometricity of General Relativity}

\author{Eleanor Knox\\New College\\Oxford University\\Oxford\\OX1 3BN\\UK\\ \texttt{eleanor.knox@new.ox.ac.uk}}
\date{\today}

\maketitle
\begin{abstract}

I look at the `flavour-oscillation clocks' proposed by D.V.
Ahluwalia, and two arguments of his suggesting that such clocks
might behave in a way that threatens the geometricity of general
relativity (GR). The first argument states that the behaviour of
these clocks in the vicinity of a rotating gravitational source
implies a non-geometric element of gravity. I argue that the
phenomenon is best seen as an instance of violation of the `clock
hypothesis', and therefore does not threaten the geometrical
nature of gravitation. Ahluwalia's second argument, for the
`incompleteness' of general relativity, involves the idea that
flavour-oscillation clocks can detect constant gravitational
potentials. I argue that the purported
`incompleteness-establishing' result is in fact one that applies
to all clocks. It is entirely derivable from GR, does not result
in the observability of the potential, and is not at odds with any
of GR's foundations.

\end{abstract}

\section*{Introduction}

What makes general relativity a `geometrical' theory? The
answer surely exceeds the scope of a single paper. To give
necessary and sufficient conditions for `geometricity' would at
the very least presuppose far greater consensus and clarity
concerning the concept of geometry than in fact exists in the
literature. Nonetheless, it is certainly possible to highlight
aspects of GR that have lead to the popular view that it is a
theory of spacetime geometry. Einstein's famous insight, that many
of the effects that have traditionally been conceived of as
gravitational may in fact be thought of as consequences of
acceleration relative to a local inertial frame, is clearly
essential to the geometric nature of general relativity. However,
despite the uncontroversial nature of this claim, the importance
of a clear understanding of this idea is sometimes ignored,
obscuring what little clarity exists in the foundations of general
relativity.

Ahluwalia \cite{Ahluwalia:1997} and Ahluwalia and Burgard
\cite{Ahluwalia:1996} have put forward a model for a flavour
oscillation clock, providing an interesting investigation of
gravitational effects on quantum systems. In two papers
\cite{Ahluwalia:1997, Ahluwalia:1998} investigating the
behaviour of such clocks in gravitational fields, Ahluwalia has
suggested that these clocks challenge the idea that general
relativity might provide a good model for gravity in the quantum
realm. On the one hand, the failure of the clocks to red-shift in
accordance with GR demonstrates a non-geometric aspect of the
theory, and on the other hand, their sensitivity to a constant
gravitational potential renders general relativity incomplete. I
argue that neither of these arguments go through; neither effect
is at odds with the foundations of GR. Moreover, there is nothing peculiarly quantum about either effect; both have analogs within the non-quantum domain.

In section 1, I look at Ahluwalia's claim that certain such clocks
behave oddly in the presence of a rotating mass; they fail to
redshift as general relativity predicts. This leads him to the
conclusion that he has discovered a `non-geometric element' in
gravity. I argue that a failure to fully acknowledge that
gravitational redshift is a consequence of acceleration leads
Ahluwalia to overstate the case. The anomalous redshift of certain
flavour-oscillation clocks can be reproduced by accelerating the
clock appropriately. This implies that such clocks violate the
\emph{Clock Hypothesis}, which states that the ability of a clock
to read the proper time along its worldline should be unaffected
by its state of motion. While this clock hypothesis violation may
be interesting in its own right, it does not itself threaten the
geometrical character of gravitation.

I examine Ahluwalia's argument for the `incompleteness' of a
general relativistic description of gravitation in section 2. The
suggestion here is that the rate of ticking of a flavour
oscillation clock will be influenced by a constant gravitational
potential. This implies that the rate of a clock in free-fall in
the gravitational field of the earth will fail to match the rate
of a clock in flat space at spatial infinity; Ahluwalia
claims this to be a direct contradiction of the tenets of GR. I will
argue that \emph{any} clock is, in a sense, `sensitive', to a gravitational potential, but the effect is not measurable and the result is derivable within a GR framework.

Throughout the paper, I will appeal to the Einstein Equivalence
Principle. This has been stated in many forms, often with subtly
different implications. However, for the purposes of this
discussion, I will use the following definition:

\newtheorem*{equivalence}{EEP}

\begin{equivalence}
Locally (i.e. on a scale in which tidal forces are not
relevant\footnote{The question `how local is local enough?' is a
tricky one, and highly context-dependent. For our purposes here,
it suffices to note that nothing in our examples depends on second
derivatives of the metric.}), no experiment can distinguish
between the effects of a gravitational field, and the effects of
acceleration.
\end{equivalence}

In the context of GR, the above might seem to presuppose a misleading distinction between inertial and gravitational fields. However, in the context of a discussion concerning possible threats to GR's geometrical status, it is useful to return to the stepping stone which established the identity of gravity and inertia, and reexamine its validity in the light of Ahluwalia's claims.

\section{Flavour-Oscillation Clocks and the Clock Hypothesis}
\subsection{Two types of flavour-oscillation clock}

The possibility of a flavour-oscillation clock rests on the fact
that  relativistic quantum mechanics allows certain particles to
exist in a superposition of mass and spin eigenstates. If a system
is constructed such that it oscillates between two such
superpositions, a kind of quantum clock is produced. Ahluwalia
suggests weak-flavour neutrinos and neutral kaons as particularly
appropriate for this set-up.

Now consider the following systems.\footnote{The account here is a
brief, informal sketch. For details, see \citep{Ahluwalia:1997} and
\citep{Ahluwalia:1996}.} For simplicity's sake, assume that a
particle has only two possible mass eigenstates (with eigenvalues
$m_1$ and $m_2$ where $m_1\neq m_2$) of spin-$\frac{1}{2}$.
Picking some axis, $z$, along which we shall measure spin, we
construct one clock such that it oscillates between states where
all spins have the same relative orientations, and a second that
oscillates between states which are superpositions of spin up and
spin down states.

Clock 1:

Oscillates between states $|Q_a\rangle$ and $|Q_b\rangle$,
where:\footnote{I have here set Ahluwalia's `neutrino mixing angle'
equal to $\frac{\pi}{4}$. Nothing in the discussion here depends
on this choice.}
\begin{equation}
|Q_a\rangle=\frac{1}{\sqrt{2}}|m_1,\uparrow\rangle+\frac{1}{\sqrt{2}}|m_2,\uparrow\rangle,
\end{equation}

\begin{equation}
|Q_b\rangle=-\frac{1}{\sqrt{2}}|m_1,\uparrow\rangle+\frac{1}{\sqrt{2}}|m_2,\uparrow\rangle.
\end{equation}

Clock 2:

Oscillates between states $|Q_c\rangle$ and $|Q_d\rangle$, where:
\begin{equation}
|Q_c\rangle=\frac{1}{\sqrt{2}}|m_1,\uparrow\rangle+\frac{1}{\sqrt{2}}|m_2,\downarrow\rangle,
\end{equation}

\begin{equation}
|Q_d\rangle=-\frac{1}{\sqrt{2}}|m_1,\uparrow\rangle+\frac{1}{\sqrt{2}}|m_2,\downarrow\rangle.
\end{equation}

\noindent In the absence of gravity, the probability of a
transition from state $|Q_a\rangle$ to $|Q_b\rangle$ is the same
as the probability of a transition from state $|Q_c\rangle$ to
$|Q_d\rangle$, and thus the two clocks tick at the same rate.

If we place the clocks at some fixed distance from a non-rotating,
weak gravitational source, the two mass eigenstates each pick up a
\emph{different} relative phase due to the approximately newtonian
potential. This causes both clocks to redshift identically by the
amount predicted by general relativity.\footnote{For details of
these relative phases and the effective redshift, see
\citep[p.1496-1497]{Ahluwalia:1997}.}

However, the situation changes if we consider a situation in which
the gravitational source is rotating about the $z$-axis along which
particle spin is to be measured. In this case, the gravitational
phases depend not only on mass, but also on spin. In particular, the
phase shift experienced by each eigenstate depends on how the spin
is oriented relative to the axis of rotation of the source. In the
case of Clock 1, the additional phase shift due to the rotation is
the same for both halves of each state. Because the transition
probabilities depend only on relative, and not absolute, phase
shifts, this extra spin related phase has no effect on the rate at
which the clock ticks, and Clock 1 continues to redshift according
to the predictions of GR.

Clock 2, on the other hand, is oscillating between states
involving superpositions of opposite spin states. The two
eigenstates involved in states $|Q_c\rangle$ and $|Q_d\rangle$
will therefore pick up equal and opposite phases as a result of
the interaction of spin with the rotating source.\footnote{As will
shortly be explained, this language is strictly inaccurate in the
GR context, where rather than interacting with the source, the
clock is affected by changes to the local inertial structure.
Throughout my description of these effects, it will occasionally
be necessary to adopt Ahluwalia's conceptual framework which
treats gravity as a force, rather than a modification of spacetime
structure. I hope this will not cause too much confusion.} Clock
2, unlike Clock 1, therefore picks up an overall relative phase
in this situation. This relative phase
does affect the transition probabilities, and hence the ticking
rate of Clock 2 is altered. As a result, flavour oscillation
clocks of the second variety do not experience the standard
gravitational redshift when placed in the vicinity of a rotating
gravitational source.

\subsection{Quantum mechanics and gravity}

Before examining Ahluwalia's diagnosis of the above effect, it is worth making some comment concerning the nature of his predictions. The above discussion (and that of the second half of this paper) compare the behaviour of flavour-oscillation clocks in a gravitational field with general relativistic effects such as gravitational redshift. However, in the absence of a theory of quantum gravity, it might seem mysterious that we can discuss the matter at all. In particular, it seems odd that we can use ordinary non-relativistic quantum mechanics to predict the behaviour of the clock. The model above must therefore be seen as a particularly crude approximation to some real theory which we have yet to develop; it is tempting to assume any odd effects predicted flavour-oscillation clocks are merely a consequence of the flaws of our approximation.

Let us therefore examine the approximation used above. Ahluwalia's rough-and-ready method is in fact a tried-and-tested one: simply work in the weak field limit of GR, in which the first diagonal component of the metric ($g_{00}$) may be treated as a Newtonian potential. In the case of a non-rotating body, the metric in question is just the Schwarzschild metric, and the potential just the ordinary Newtonian one. In the case of the rotating body, we use the Kerr metric, and the potential picks up a term that depends on the angular momentum of the body. In both cases, quantum predictions are generated simply by incorporating these potentials into the Hamiltonian.

What justifies this method? It is, of course, rather natural, in the absence of a theory of quantum gravity, simply to return to a Newtonian picture, but use GR to generate predicted potentials. On the other hand, its relative naturalness is no guarantee of the accuracy of its results. Nonetheless, the method has worked in some experimentally confirmed cases; for example, the theoretical prediction of the experimentally verified COW effect \citep{COW} uses just this approximation. Moreover, the method derives some theoretical support from a result due to Rosen \cite{Rosen:1972}. This shows that non-relativistic quantum mechanics is form-invariant under arbitrary accelerative translational motion; accelerating a quantum system has just the same effect as subjecting the system to a potential that affects all parts of the system equally. That is, quantum mechanics obeys the Einstein Equivalence Principle; accelerations and gravitational fields may be modeled in exactly the same manner. If we therefore see the key component of GR as being the insight that a body subject to gravity is accelerating, the inclusion of a potential in the Hamiltonian seems well-motivated.

Inevitably, this is not quite the whole story. Rosen's result does not deal with rotating systems, so the case of the rotating mass may be less straightforward. Moreover, there is something a little odd about using non-relativistic quantum mechanics to model mass superpositions, which are only permitted when we move to the relativistic case. Nonetheless, Ahluwalia's approximation has some precedent. Ultimately, the arguments of this paper will in fact lend his model some support, by suggesting that the strange effects he discusses are not peculiar to quantum systems at all.

\subsection{A new non-geometric element in gravity?}

How are we to interpret the anomalous redshift? Ahluwalia suggests
we draw the following conclusions:

\begin{quote}
...the non-geometric element in redshifts may be interpreted as a
quantum-mechanically induced fluctuation over a geometric structure
of space-time.\citep[p.1500]{Ahluwalia:1997}
\end{quote}

Ahluwalia sees the phenomenon as a quantum fluctuation, because
there in fact exists a third variety of flavour-oscillation clock,
formed by interchanging the spin states of clock 2. In the
presence of a rotating source, this clock will experience a change
in rate exactly opposite to that experienced by clock 2. As a
result, if we take an equally weighted sample of our three types
of clock, the average shift will precisely be the gravitational
redshift predicted by GR.\footnote{Of course this isn't a quantum
fluctuation in the usual sense, as there is no uncertainty as to
which type of clock some individual system comprises.}

Leaving the issue of fluctuations aside, why claim that these
redshifts possess a `non-geometric' character? For Ahluwalia, this
claim depends on the following definition:

\newtheorem{geometry}{Geometry}

\begin{geometry}\label{geom1}
Geometrical elements are those that are completely specified by the
gravitational source. Non-geometrical elements are those that
crucially depend on the details of quantum test particles and do not
follow from general relativity alone. \citep[p.1494]{Ahluwalia:1997}
\end{geometry}

The idea expressed here is understandable. The universality of
gravity is central to its amenability to geometrical
interpretations. However, it is far from certain that the above is a
useful way to cash out the notion of universality that lies at the
heart of general relativity's geometrical nature.

For comparison's sake, let us turn briefly to another effect
arising from a combination of gravity and quantum mechanics: the
neutron interferometry results from Colella, Overhauser and
Werner's famous experiment \citep{COW}. In this setup, the two
halves of a neutron beam split in a tilted crystal interferometer
pick up a phase difference as a result of the (minutely) varying
gravitational potential experienced by each of the two beams. The
resulting phase shift turns out to depend on neutron mass. It is
possible to view this as a violation of the spirit of general
relativity. Such a result constitutes a `non-geometrical element'
by Ahluwalia's lights; the magnitude of the effect depends not
only on the gravitational force but also on the details of the
quantum test particle.

However, there is good evidence that the COW experiment, despite
appearances, is not actually at odds with GR. In particular,
despite the appearance of a mass term in the phase shift, there
need be no suggestion that the experiment violates the Einstein
Equivalence Principle. Exactly the same result is predicted if the
apparatus is accelerated, a prediction that has been borne out by
experiments involving horizontally accelerated interferometers
\citep{Bonse:1983}. It is therefore natural to see the effect as
resulting from the acceleration of the interferometer with respect
to freely falling inertial frames. Assuming that the equivalence of gravity and inertia is central to what it means for General Relativity to be a
geometrical theory, it is hard to make precise the notion that the
COW experiment demonstrates a `non-geometrical'
effect.\footnote{For further details of the COW experiment, and an
overview of other neutron interferometry results, see
\citep{Greenberger:1983} and \citep{Werner:1994}.}

It would therefore seem to make sense to ask the same question with
regard to Ahluwalia's anomalous redshift: Is it possible to view
this effect as purely a result of the acceleration of the clocks
relative to local inertial frames? We will see in the next section
that the answer is yes.

\subsection{Cancelation and Simulation: An alternative account of
geometricity.}

Chryssomalakos and Sudarsky \cite{Chryssomalakos:2003} have argued
that the behaviour of neutrino-oscillation clocks does not in fact
impugn the geometrical nature of gravitation. In the course of their
argument, they introduce the following approximate
`phenomenological' criterion for geometricity:

\begin{geometry}\label{geom2}
Gravity is geometrical if all its effects can be locally canceled
(or simulated) by a suitable choice of reference frame in which
their description takes place. \citep[p.607]{Chryssomalakos:2003}
\end{geometry}

As it stands, this definition is unsatisfactory. It is most
certainly not the case that all interesting gravitational effects
can be canceled simply by altering the reference frame from which
they are described: if Clocks 1 and 2 are comoving, and fail to
tick at the same rate, there is no choice of frame which can
negate the effect. However, Chryssomalakos and Sudarsky
are clearly trying to capture the insight that springs from the equivalence principle, that gravitation and inertia are, in the context of GR, two sides of the same coin.

Bearing in mind that, once we move to the full GR picture, the language of simulation and cancelation is not a natural one, we may note that one consequence of the Einstein Equivalence Principle as defined in the previous chapter is that the effects of a
homogeneous gravitational field on a system can be simulated, in
the absence of gravity, by accelerating it. Another consequence is
that a freely falling object should behave as a force-free body
would. If we take these two notions instead of
Chyssomalakos and Sudarsky's notions of simulation and cancelation
in {\bf Geometry 2}, then we appear to have a relevant and
pragmatic criterion for the `geometricity' of a phenomenon:

\begin{geometry}
Gravity is geometrical only if its (local, non-tidal) effects
disappear entirely when an object is freely falling, and its effects
can be simulated by accelerating a system appropriately.
\end{geometry}

I have phrased this as a necessary, rather than sufficient
condition because the subtleties of general relativity's
geometrical character require further examination.  Nonetheless, it is interesting and relevant to examine
whether Ahluwalia's flavour-oscillation clocks violate {\bf
Geometry 3}.

Chryssomalakos and Sudarsky consider the effects of viewing Clock 2
from a local inertial (free-fall) frame. They then note that from
this perspective, the difference in the rates of Clock 1 and Clock 2
will persist, but if the clocks are themselves put into free-fall,
the effect disappears:

\begin{quote}
...one might conclude that the effect would persist in the freely
falling frame. This would be very puzzling to say the least.
However, we must be careful and note that if all we do is change the
frame of description but not make the experimental apparatus
(including the detectors) move with the locally inertial frame, then
the above-mentioned situation would ensue. On the other hand, if we
make the experimental apparatus (in particular, the detectors) move
together with the locally inertial frame, then the effect will
indeed disappear as it should. \citep[p.613]{Chryssomalakos:2003}
\end{quote}

Thus Chyssomalakos and Sudarsky suggest that the particular
gravitational effect in question does disappear in free-fall. As
this is the result on which they base their argument for the
geometricity of gravity, it is quite clear that the criterion that
they have in mind is {\bf Geometry 3} rather than {\bf Geometry
2}. Their concern is not to show that the difference in ticking
rate of clocks 1 and 2 may be canceled by moving to an appropriate
frame, but rather that the effect Ahluwalia discusses disappears
in free-fall.

What of the issue of simulation? Is there an accelerative
situation in which we can expect Clocks 1 and 2 to display the
same rate difference as they do in the presence of a rotating
gravitational source? With some thought, it becomes clear that
there is. In GR, it is well-known that a rotating mass `drags' the
local inertial frames in such a way that they are themselves
rotating relative to inertial frames at infinity. This is the
Lense-Thirring effect, first predicted in 1918
\citep{Lense-Thirring}.\footnote{ See also \citep{Brill:1966}. For a
straightforward overview of gyroscope precession and the
construction of an inertial frame based on gyroscopes, see
\citep{Hartle:2003}.} It therefore follows that a body held
stationary above such a rotating mass is not only accelerating,
but also rotating relative to the local inertial frames. This
suggests that the effect can be easily simulated: simply put
Clocks 1 and 2 into the appropriate accelerated, rotating motion.
In such a situation, the spin states of the two clocks would be
expected to be affected by their rotational motion in exactly the
same way that they are affected by the rotating gravitational
source. We therefore have no reason to believe that
flavour-oscillation clocks produce any effects at odds with {\bf
Geometry 3}.

\subsection{The Clock Hypothesis}

It might be countered that my comments above do not
rule out Ahluwalia's conclusion. I have explicitly said that {\bf
Geometry 3} is a necessary, rather than sufficient, criterion for
geometricity. This certainly leaves logical space for the
insistence that Ahluwalia's {\bf Geometry 1} is also a necessary
condition for geometricity, and that, by its lights, the
flavour-oscillation clock result is `non-geometrical'.

Nonetheless, I think a clear understanding of the Einstein
Equivalence Principle and {\bf Geometry 3} erodes the ground for
such a position. For, once we understand that the results of
Ahluwalia's thought experiment would be replicated by simply
putting the clocks into accelerated, rotational motion, we should,
in line with a standard reading of GR, see the clocks in a
vicinity of a rotating source as simply \emph{in} such
accelerated, rotational motion relative to the local inertial
frames. On such a view, Ahluwalia's second variety of
flavour-oscillation clock is simply a clock that is affected by
certain kinds of motion. That is, we can see it as a clock that
fails to obey the \emph{Clock Hypothesis}:\footnote{ The clock
hypothesis is not often discussed in detail in the literature. A
notable exception is \citep[ p.94-95]{Brown:2005}. For an account of
the behaviour of atomic clocks under acceleration, see
\citep{Ohanian:1994}, p.172.}

\newtheorem*{clock hypothesis}{Clock Hypothesis}

\begin{clock hypothesis}
The rate of a clock depends only on its instantaneous velocity:
\begin{equation}
\Delta{t}=\int^{t_2}_{t_1}(1-\frac{v^2}{c^2})^{\frac{1}{2}}dt=\int^{t_2}_{t_1}d\tau,
\end{equation}
\n where $\Delta{t}$ is the time read by the clock, and $\tau$ is
proper time.\footnote{I have written the clock hypothesis here as it is usually stated. However, Ahluwalia's predictions for his clocks assume low velocities, and thus the $\frac{v^2}{c^2}$ term above is in fact irrelevant in the current case.}
\end{clock hypothesis}

This amounts to asserting that clocks will read off the lengths of
their world-lines even when their path is not a geodesic. The
validity of the clock hypothesis rests on the idea that a good
clock's mechanism should not be influenced by accelerations. For a
standard, macroscopic clock, this amounts to something like the
assumption that the internal restorative forces which ensure the
function of the clock are large compared to the forces that
accelerate it. Clearly, while this assumption is reasonable under
a wide range of circumstances, it will inevitably fail for certain
clocks under certain conditions. Standard mechanical clocks would
suffer if we hit them with a hard enough force, and that most
classic and reliable of clocks, the pendulum, ceases to work when
subjected to any significant acceleration whatsoever.\footnote{Of
course, from the perspective of GR, we should say that a pendulum
\emph{only} works when subjected to a constant acceleration.}

Under these circumstances, why should we regard
flavour-oscillation clocks as particularly inimical to the
geometricity of GR? Why is a clock that happens to be affected by
rotational motion for well-understood reasons (its ticking rate is
intimately connected to spin eigenstates) any more disturbing than
a clock that ceases to work when we hit it?\footnote{ Of course,
the fact that one case involves rotational and the other linear
acceleration could be said to constitute a difference in itself,
but not one that is relevant here; both cases involve
well-understood failures of the clock hypothesis. Nonetheless, the
failure of quantum bodies with spin to behave as they should in
General Relativity is interesting, and may occur in other cases.
Drummond and Hathrell \cite{Drummond:1980} have calculated the
effective action in QED with a gravitational field, and found that
it predicts deviation from the null geodesics for photons in
certain circumstances. In many ways, this effect provides a more
serious threat to the geometricity of GR than flavour-oscillations
clocks, because the effect involves a violation of minimal
coupling. A thorough discussion of the effect may be found in
\citep[p.165-168]{Brown:2005}.}

Consider, for example, comparing a rotating type 2
flavour-oscillation clock with a standard, old-fashioned,
mechanical pocket watch undergoing extreme linear acceleration. In
both cases, we have a clear understanding of why certain types of
motion affect the workings of the clock. In the case of the watch,
its ticking depends on the ability of its spring-balance to
operate. Subject the watch to a force that is large compared to
the force exerted by the spring balance, and it will cease to
work. In the case of the flavour-oscillation clock, its ticking
depends on the relative phase between two spin eigenstates.
Subject the clock to a rotational acceleration, and the spin of
the two eigenstates will couple to the clock's angular momentum,
altering the rate of the clock. The latter case seems no more
mysterious than the former. Certainly, it is clear that the problem does not arise as the result of any peculiarities arising from the \emph{quantum} nature of the flavour-oscillation clock; any number of non-quantum clocks would be affected by rotational motion.

In the light of this, it appears that we have two choices. First,
we can accept that flavour-oscillation clocks do not, in fact,
threaten the geometricity of gravitation, and that {\bf Geometry
1} is not a plausible criterion. Second, we can accept {\bf
Geometry 1}, but then accept that the sensitivity of my
wrist-watch to accelerations had long-since undermined any
possibility of interpreting GR as a geometrical
theory.\footnote{Of course, Ahluwalia's {\bf Geometry 1} makes
explicit reference to a quantum test-particle, but the general
thrust is that the behaviour of a clock must not depend on its
constitution.} I feel confident that most readers would choose the
former.

\section{Flavour-Oscillation Clocks in a Constant Potential}

In spite of the above, further threats to the geometricity of
gravity loom. In a second paper \citep{Ahluwalia:1998}, Ahluwalia
makes an even stronger claim for his flavour-oscillation clocks:
not only are they held to demonstrate a non-geometric aspect of
gravitation, but their behaviour in a constant gravitational
potential demonstrates the ``incompleteness'' of general
relativity. According to Ahluwalia, General Relativity requires
that a clock in free-fall tick at the same rate as an inertial
clock at spatial infinity in an asymptotically flat spacetime.
Flavour-oscillation clocks fail to adhere to this, and thus
produce a result at odds with the foundations of GR. In section 2.1 I will attempt to reconstruct Ahluwalia's reasoning. I will then argue in section 2.2 that, for a number of reasons, the argument does not go through; the behaviour of flavour-oscillation clocks is not at odds with GR.

\subsection{The problem according to Ahluwalia}

According to Ahluwalia:

\begin{quote} The conceptual basis of the
theory of general relativity asserts that the flat space-time
metric...is measured by a freely falling observer on Earth (or,
wherever the observer is).\citep[p.4]{Ahluwalia:1998}
\end{quote}

As a result, claims Ahluwalia, a clock in freefall should tick at
the same rate as a clock at spatial infinity. However, the rate of
a flavour-oscillation clock in free-fall is sensitive to a
constant potential, and thus will not be the same as the rate of a
clock in flat Minkowski spacetime. Let us see how the argument
works:

Consider a flavour oscillation clock of the type introduced in
section 1.1. For the purposes of this argument, the spin
eigenstates play no significant role, so it is irrelevant whether
we consider clocks of type 1 or 2. There are two possible ways of predicting the effects of gravity on a quantum system. One option is to solve the the Schr\"odinger equation with
a gravitational interaction energy term, treating the gravitational potential as a potential like any other:

\begin{equation}
\left[-\left(\frac{\hbar^2}{2m_i}\right)\nabla^2+m_gV_{grav}(\mathbf{r})\right]\psi(t,\mathbf{r})=i\hbar\frac{\partial\psi(t,\mathbf{r})}{\partial
t}.
\end{equation}

Alternatively, it is possible to take a route that respects general relativity's geometrical view of gravity.  As the work of Stodolsky \cite{Stodolsky:1979} has demonstrated, a particle traversing a classical path acquires a phase $\phi$ that depends on its mass and its proper time:
 \begin{equation}
\Phi=\int_A^Bmds.
\end{equation}
\n In the weak-field limit for slow moving particles, this becomes:
\begin{equation}
\Phi=\Phi_{0}+\phi,
\end{equation}
where $\Phi_{0}$ is the phase picked up when no potential is present (at non-relativistic velocities):
\begin{equation}
\Phi_{0}=mt,
\end{equation}
and $\phi$ is the gravitational part of the phase:
\begin{equation}
\phi=mV_{grav}t.
\end{equation}
\n This result accords with the solution to the above Schr\"odinger equation with a classical potential. By calculating the probability of a transition from state $|Q_a\rangle$ to state $|Q_b\rangle$, we find that the rate of ticking when the potential is zero is given by:

\begin{equation}
\Omega_{0}=\frac{(m_2-m_1)c^2}{2\hbar}.
\end{equation}

\n If we now place the clock into some gravitational potential, we find that the ticking rate picks up an additional factor due to the change in relative phase caused by the gravitational phase:
\begin{equation}
\Omega_{V_{grav}}=\left(1+V_{grav}\right)\Omega_{0}.
\end{equation}

Ahluwalia now considers the case of a spherically symmetric, weak field in general relativity, such as that surrounding the earth. In this situation, the metric is given by:

\begin{equation}
ds^2=\left(1+2V_{grav}(\mathbf{r})\right)dt^2-\left(1-{2V_{grav}(\mathbf{r})}\right)d\mathbf{r}^2.
\end{equation}

\n A flavour-oscillation clock held stationary above the earth's surface experiences a Newtonian potential $-\frac{GM_E}{r}$. Putting this into equation 12, we get the standard gravitational redshift.

Thus far everything is in accordance with GR, but Ahluwalia has another worry. Equation 12 is not invariant under the addition of a constant gravitational potential. Ahluwalia uses this fact to generate an apparent contradiction with GR. There exists a distant but massive galactic cluster called the Great Attractor. According to
Newtonian gravitational theory, the
gradient of the potential may be ignored because it drops off as the square of the distance. However, the potential itself has a comparatively large
numerical value. For all intents and purposes, Ahluwalia claims, we may consider this potential to be constant. However, a constant potential does not affect the gravitational acceleration of the clock, which depends solely on the gradient of the potential. Because ticking rate is affected by such constant additions to the potential, all neutrino oscillation clocks on
earth should be affected by the great attractor. Moreover \emph{even clocks in freefall} will be affected: a constant potential does not influence gravitational acceleration, and therefore is not cancelled when we transform to the freefall frame. When a clock is put into freefall in some potential $V=-\frac{GM_E}{r}+V_{const}$, it should tick at the following rate:

\begin{equation}
\Omega_{freefall}=\left(1+{V_{const}}\right)\Omega_{0}.
\end{equation}.

This, Ahluwalia claims, is a direct violation of the foundations of GR, which he takes to imply that a clock in freefall must tick at the same rate as a clock at spatial infinity (i.e., in flat spacetime). Assuming that the potential at spatial infinity is zero, this appears not to be the case. General relativity is, Ahluwalia claims, ``incomplete".

\subsection{Unpicking the argument}

The above argument possesses many strands, and some untangling and reconstruction will be necessary in what follows. Ultimately, even when the problem is put in a more plausible form, we will see that it does not go through.

First, let us turn to Ahluwalia's account of the foundations of GR with which his effect is claimed to be at odds. Recall, there are two related claims here: First, that a freely falling observer should measure the Minkowski metric, and second, that a clock in freefall should tick `at the same rate' as one at spatial infinity.

The first of these is clearly false. GR does not demand that
freely falling observers measure the Minkowski metric. What GR does demand is that, in as much as a freefall
observer may be considered to occupy a local inertial frame (i.e. as long as tidal effects may be ignored), she is an inertial observer; that she follows a geodesic, and
feels no gravitational forces.
The Einstein Equivalence Principle further demands that no
\emph{local}\footnote{Here, local simply means on any scale at
which tidal effects are not relevant.} observations allow the
freefall observer to distinguish between herself and an inertial
observer in flat Minkowski spacetime, but this should not be
extrapolated to the conclusion that any freefall observer measures
the Minkowski metric. \emph{Locally}, any observer, free-fall or
otherwise, will measure the Minkowski metric - the structure of GR
spacetime is locally Minkowskian - but \emph{globally}, all
observers will agree on whether the spacetime they are in is flat
or curved.

As for the second, it is not entirely clear what comparing the rate of a free-fall clock to a clock at spatial infinity is supposed to represent. It is not possible to directly compare the rates of two distant clocks; we do not peer across space and determine their relative rates. What we may do is compare the rate of one clock to the time coordinates most naturally defined by the other. Operationally, such a procedure corresponds to comparing the time elapsed on one clock with the difference in time between two appropriately synchronised clocks at rest with respect to one another, as our first clock whizzes past the other two. This is what we mean when we refer to the `difference in ticking rate' represented by relativistic time dilation. Is this kind of procedure what Ahluwalia has in mind? If it is, then the claim that a clock in freefall must tick at the same rate as one at infinity cannot be right. Both clocks may, according to GR, be inertial clocks, but, even in flat spacetime, inertial clocks do not tick `at the same rate' in the above sense. In Minkowski spacetime, one inertial clock will generally experience time dilation relative to the coordinate time defined by two comoving, synchronised, clocks.

Thus, the very foundations of GR with which Ahluwalia claims his results to be at odds turn out not to be foundations of GR at all. Does this close the issue? Not quite; we may still use Ahluwalia's results to derive a problem. In GR, as in Newtonian mechanics, the absolute value of the gravitational potential should not be measurable. This is particularly clear in GR, where gravitational structure \emph{is} inertial structure. The addition of a constant potential clearly does not influence inertial structure, and ought, therefore, have no measurable effects. Another way of seeing this is to note that, in the weak field limit where the concept of a Newtonian potential may be applied, neither the connection coefficients nor the Riemann curvature are affected by the addition of a constant potential to the diagonal terms of the metric; both depend only on derivatives of the metric components and not their absolute values. GR therefore requires that, if there are cases in which it is permissible to add a constant term to the potential in the Newtonian limit, the effect of such a term must not be measurable.

Given the above, we might think that Ahluwalia's result \emph{is} at odds with GR. First, there is claimed to be some physical situation in which a constant gravitational potential ought to be inserted into the metric. Second, it is claimed that such a constant potential will affect the rate of a certain kind of clock. Prima facie, this is at odds with GR, at least if the effect turns out to be measurable.

We will shortly see that this effect will not in fact be measurable. However, let us first turn to what justifies the introduction of a constant potential in the first place. The metric given in equation 13, with $V=-\frac{GM}{r}$, is derived from the Schwarzschild solution to the GR field equations, with the addition of the following assumptions: First, we assume that the field is sufficiently weak that we may apply the linearised form of the metric:

\begin{equation}
g_{\mu\nu}=\eta_{\mu\nu}+h_{\mu\nu},
\end{equation}

\n with $|h_{\mu\nu}|\ll1$. However, this $h_{\mu\nu}$ is not unique; the form of the above is retained either under global Lorentz transformations, or under infinitesimal coordinate transformations. In order to simplify the field equations, we work in the Lorentz gauge. That is, we assume the following condition:
 \begin{equation}
\bar{h}\indices{^{\mu\alpha}_{,\alpha}}=0
 \end{equation}
 \n where
 \begin{equation}
\bar{h}_{\mu\nu} \equiv h_{\mu\nu}-\frac{1}{2}\eta_{\mu\nu}h.
 \end{equation}
 However, this still does not uniquely fix the value of $h_{\mu\nu}$. Ordinarily, we also fix the above such that $h_{\mu\nu}(r=\infty)=0$, so that our coordinates are Lorentzian far from the source. This ensures that $h_{00}=h_{ii}=2V$, where $V$ is the Newtonian potential which is also zero at infinity. However, this final condition is not implied by any of the equations of motion, which only fix the gradient of the potential, and not the potential itself. In this case, we may, if we wish, choose not to work with the coordinates imposed above, and adopt some potential $V=-\frac{GM}{r}+V_{const}$, accepting, of course, that our potential will now be non-zero at infinity. This choice of coordinates\footnote{The exact meaning of the alternative coordinate choice will be examined shortly.} should, however, have no measurable effects; the equations of motion are invariant under the addition of a constant term to $h_{\mu\nu}$.

The above gives a simple example of a case in which the introduction of a constant potential is allowed. However, it is worth noting that this situation is very different from the one that Ahluwalia describes, despite the fact that he uses a metric that precisely corresponds to the weak field schwarzschild metric expressed in a non-standard gauge:

\begin{equation}
ds^2=\left(1-2(\frac{GM_E}{r}+V_{const})\right)dt^2-\left(1+2(\frac{GM_E}{r}+V_{const})\right)d\mathbf{r}^2.
\end{equation}

In Ahluwalia's case,  $V_{const}$ is the potential generated by the great attractor. However, this means that the above is strictly inaccurate. The gradient of the potential due to the great attractor is not zero, but merely very small. As a result, the earth, along with any flavour oscillation clocks in its vicinity, will be freely falling towards the great attractor. Just as the earth's gravitational potential disappears when viewed from the rest frame of a clock falling freely towards it, so too will the potential of the great attractor disappear as the flavour oscillation clock falls freely towards it.

Ahluwalia's example is therefore deeply misleading, not least because it maintains the illusion that we can introduce a constant potential that also falls to zero at infinity! Nonetheless, there are other weak-field situations in which the addition of a constant potential may be appropriate; for example, in the simple spherically symmetric case described above. As a result, if it were the case that flavour-oscillation clocks could detect the absolute value of the potential, GR would still be in trouble.

So, is the effect predicted by Ahluwalia measurable? If it were the case that \emph{only} flavour oscillation clocks were affected by the addition of a constant potential, it most certainly would be. The ratio of flavour oscillation clocks to other clocks would enable the measurement of the potential.\footnote{It is not clear to what extent Ahluwalia believes that his effect applies to flavour-oscillation clocks alone. At one stage \citep[p.7]{Ahluwalia:1998}, he seems to suggest that all terrestrial clocks will be affected. At another, \citep[p.3]{Ahluwalia:1998} he implies that only certain quantum systems are sensitive to the potential. At any rate, he fails to see that the effect is both unmeasurable, and, in a certain sense, a straightforward consequence of GR itself.} Fortunately for GR, the effect is, in fact, universal. \emph{Any} clock is `sensitive' to the value of the potential. This was first pointed out by Stodolsky \cite{Stodolsky:1979}, who considers the possibility of measuring the gravitational potential by means of a clock based on a mass superposition:

\begin{quote}
The $K^0$ clock...slows down when we put it in a region of smaller
$g_{00}$, but then so do all other clocks-the red shift is universal...[W]e are rescued from observing the value of $g_{00}$ locally by a general coordinate transformation which slows down all clocks
equally.\citep[p.395]{Stodolsky:1979}
\end{quote}

We do not need to look at the details of a quantum clock to derive the effect of a potential on ticking rate; it is a consequence of general relativity itself. Consider a clock at rest in the coordinate system in which the weak Schwarzschild metric takes the form given by equation 14. Relative to this frame our clock is at rest, so $d\mathbf{r}=0$, and we have:

\begin{equation}
ds^2=\left(1+2(\frac{GM_E}{r}+V_{const})\right)dt^2.
\end{equation}

\n For small potentials, this gives:

\begin{equation}
\Omega_{V_{const}}=\left(1+V_{const})\right)\Omega_{V_{const}=0},
\end{equation}

\n where $\Omega_{V_{const}=0}$ is the rate the clock ``would have ticked at", if the constant part of the potential had been zero. The clock appears to be `affected' by the constant potential. This is just another form of Ahluwalia's result, derived straight from GR without any details of flavour-oscillation clocks at all! It appears that Ahluwalia's effect is in fact universal. However, it is worth noting that while we remain in the non-standard gauge\footnote{ I here use the term `standard gauge' to refer to the situation in which we impose the vanishing of the potential at infinity, and non-standard gauge to indicate that this is not the case.} for which the addition of a constant potential has some meaning, there is no region of space for which this potential is zero. The only way to make sense of zero potential, and hence of the comparison to $\Omega_{V_{const}=0}$ embodied by the above, is to take $\Omega_{V_{const}=0}$ to be the ticking rate as described in the standard gauge, in which the potential is zero at infinity. There is no way to have a constant potential in some places and zero potential in others while we remain in a single gauge.

This observation leads us into philosophical waters. What should we make of such comparisons of ticking rate across solutions with different boundary conditions? How are the two solutions related? As a result of the diffeomorphism invariance of the full theory, the equations of motion of weak-field GR are invariant under gauge transformations of the weak field potential:
\begin{equation}
h_{\mu\nu}\rightarrow h_{\mu\nu}+\xi_{\mu,\nu}+\xi_{\nu,\mu}
\end{equation}
\noindent where $\xi$ are functions small enough to leave the transformed potential small.
Our two weak field solutions are related by just such a transformation; we go from one to the other simply by adding a constant term to $h_{\mu\nu}$. We can therefore see our freedom to add a constant potential as a simple instance of the diffeomorphism invariance of GR. What transformation have we imposed? Adding a constant potential everywhere amounts to scaling all time units by a factor of $(1+V_{const})$, and scaling distance units by a factor of $(1-V_{const})$. This makes sense of the idea that, relative to the old coordinates, all clocks in our new gauge tick at a different rate. We should therefore insist that the apparent changes caused by the universal addition of a constant potential do not correspond to \emph{physical} changes, but rather to mere redescriptions of the same physical situation.

Of course, this is not to deny that clocks experiencing \emph{different} potentials as expressed in a single gauge may be said to tick at different rates; this is the standard gravitational redshift. However, such comparisons depend only on potential differences, and not on the potential itself.

Thus we see that the effect discussed by Ahluwalia is universal, and we are saved from observing the value of a potential that must, in GR, be unobservable. Moreover, in the realistic case of a constant potential discussed here, the unobservable and unphysical status of the potential has a clear explanation; the addition of a constant potential just corresponds to writing the metric in terms of rescaled coordinates. It seems, therefore, that Ahluwalia's charge of incompleteness does not go through. Certainly, the application of Newtonian concepts to weak-field situations in GR generates conceptual puzzles, but none so deep that they undermine the foundations of the theory.
\section*{Conclusion}

Neither of Ahluwalia's arguments should be taken to refute general
relativity and the geometrical picture of gravity it suggests. The
fact that a clock may be constructed that is sensitive to the
rotation of a gravitational source need not, in and of itself,
threaten the geometricity of gravitation. The result can be
interpreted, within the general relativistic context, as being
caused by the rotation of the clock relative to the local inertial
frames. Certain of Ahluwalia's clocks are sensitive to rotations,
and therefore do not obey the clock hypothesis when in rotational
motion. This is interesting, but neither unique, nor uniquely quantum; there are other, classical,
instances of clock hypothesis violation. The fact that certain
systems, which function well as clocks in certain states of
motion, fail to measure proper time in all circumstances is not at
odds with the foundations of GR.

Similarly, Ahluwalia's second result, concerning the behaviour of
flavour oscillation clocks in a constant potential, does not, on
close inspection, threaten the geometricity, or the completeness,
of GR. In fact the effect discussed does
not depend on any feature special to flavour-oscillation clocks,
but is rather derivable for any clock, quantum or not. The
apparent contradiction of the tenets of GR stems, in part, from a misreading
of the consequences of the equivalence principle, proving once
more that the importance of clarity concerning the deceptively
simple foundations of GR should not be underestimated.

\bibliographystyle{plain}
\bibliography{generalbib}

\begin{thebibliography}{10}

\bibitem{Ahluwalia:1997}
D.V. Ahluwalia.
\newblock On a new non-geometric element in gravity.
\newblock {\em Gen. Rel. and Grav.}, 29(12):1491--1501, 1997.

\bibitem{Ahluwalia:1998}
D.V. Ahluwalia.
\newblock Can general-relativistic description of gravitation be considered
  complete.
\newblock {\em Modern Physics Letters A}, 13:1393--1400, 1998.

\bibitem{Ahluwalia:1996}
D.V Ahluwalia and C.~Burgard.
\newblock Gravitationally induced neutrino-oscillation phases.
\newblock {\em Gen. Rel. and Grav.}, 28:1161--1170, 1996.

\bibitem{Bonse:1983}
Ulrich Bonse and Thomas Worblewski.
\newblock Measurement of neutron quantum interference in noninertial frames.
\newblock {\em Phys. Rev. Lett.}, 51(16):1401--1404, 1983.

\bibitem{Brill:1966}
Dieter~R. Brill and Jeffrey~M. Cohen.
\newblock Rotating masses and their effect on inertial frames.
\newblock {\em Phys. Rev.}, 143, 1966.

\bibitem{Brown:2005}
Harvey~R. Brown.
\newblock {\em Physical Relativity: Space-time structure from a dynamical
  perspective}.
\newblock OUP, 2005.

\bibitem{Chryssomalakos:2003}
C.~Chryssomalakos and D.~Sudarsky.
\newblock On the geometrical character of gravitation.
\newblock {\em Gen. Rel. and Grav.}, 35(4):605--617, 2003.

\bibitem{COW}
R.~Colella, A.W. Overhauser, and S.A. Werner.
\newblock Observation of gravitationally induced quantum interference.
\newblock {\em Phys. Rev. Lett.}, 34(23):1472--1474, 1975.

\bibitem{Drummond:1980}
I.T. Drummond and S.J. Hathrell.
\newblock { QED} vacuum polarization in a background gravitational field and
  its effect on the velocity of photons.
\newblock {\em Physical Review D}, 22(2):343--355, 1980.

\bibitem{Greenberger:1983}
Daniel~M. Greenberger.
\newblock The neutron interferometer as a device for illustrating the strange
  behaviour of quantum systems.
\newblock {\em Rev. Mod. Phys}, 55(4):875--905, October 1983.

\bibitem{Hartle:2003}
James~B. Hartle.
\newblock {\em Gravity: An introduction to Einstein's General Relativity}.
\newblock Addison Wesley, 2003.

\bibitem{Ohanian:1994}
H.C. Ohanian and R.~Ruffini.
\newblock {\em Gravitation and Spacetime}.
\newblock W.W. Norton and Company, 1994.

\bibitem{Rosen:1972}
G.~Rosen.
\newblock {Galilean invariance and the general covariance of nonrelativistic
  laws}.
\newblock {\em American Journal of Physics}, 40:683, 1972.

\bibitem{Stodolsky:1979}
Leo Stodolosky.
\newblock Matter and light wave interferometry in gravitational fields.
\newblock {\em General Relativity and Gravitation}, 11(6):391--405, 1979.

\bibitem{Lense-Thirring}
H.~Thirring and J~Lense.
\newblock \"{U}ber den einfluss der eigenrotation der zentralk\"orper auf die
  bewegung der planeten und maude nach der einsteinschen gravitationstheories.
\newblock {\em Phys. Z.}, 19:156--163, 1918.

\bibitem{Werner:1994}
S.A. Werner.
\newblock Gravitational, rotational and topological quantum phase shifts in
  neutron interferometry.
\newblock {\em Class. Quantum Grav.}, 11:A207--A226, 1994.

\end{thebibliography}
\end{document}